\documentclass[12pt,preprint]{aastex}

\def\ltsima{$\; \buildrel < \over \sim \;$}
\def\lsim{\lower.5ex\hbox{\ltsima}}

\begin{document}

\title{A New Method to Map Flares in Quasars}

\author{Atsunori Yonehara\altaffilmark{1,2}}
\affil{Center for Computational Physics, 
 University of Tsukuba, 1-1-1 Tennodai, Tsukuba 305-8577, Japan}

\author{Shin Mineshige}
\affil{Yukawa Institute for Theoretical Physics,  
 Kyoto University, Kitashirakawa-Oiwake-Cho, Sakyo-ku, Kyoto 606-8502, Japan}

\author{Yoh Takei}
\affil{The Institute of Space and Astronautical Science, 
 3-1-1 Yoshinodai, Sagamihara, Kanagawa 229-8510, Japan}

\author{George Chartas}
\affil{Department of Astronomy and Astrophysics, 
 Pennsylvania State University, University Park, PA 16802, USA}

\and 

\author{Edwin L. Turner}
\affil{Princeton University Observatory, 
 Peyton Hall - Ivy Lane, Princeton, NJ 08544-1001, USA }

\altaffiltext{1}{Research Fellow of the Japan Society 
 for the Promotion of Science}
\altaffiltext{2}{e-mail: yonehara@rccp.tsukuba.ac.jp}

\begin{abstract}

Recently, Chartas et al. (2001) detected a rapid X-ray flare
in the gravitationally lensed, multiple image quasar RX~J0911.4+0551.
Dramatic events, such as rapid X-ray flares, are useful in providing 
high precision measurements of the time delays between multiple images.

In this paper, we argue that there is a new possibility in 
measurements of time delays between multiple images of 
gravitationally lensed quasars; constrain the locations of putative flares 
that give rise to the intrinsic rapid variabilities of quasars. 
The realization, however, of these goals cannot be presently achieved 
due to the limited accuracy of the current measurements. 
We predict that timing flares with accuracies of the order of 
a few seconds will be needed to probe the location of the flares.
Our proposing method will work with better instruments in near future, 
such as XEUS.

\end{abstract}


\section{INTRODUCTION}

It is well known that gravitationally lensed quasars 
with multiple image and their flux variabilities provide us 
oppotunities to measuring $H_{\rm 0}$ (Refsdal 1964).
After the first detection of the lensed quasar Q0957+561 
by Walsh, Carswell, \& Weymann (1979), 
surveys in the optical and radio bands have uncovered 
over seventy gravitationally lensed quasars.
For the purpose of meauring $H_{\rm 0}$, monitoring of 
several of these lensed objects has performed, 
and resulted in measurements of time-delays in about eight systems 
with accuracies of about 10\%.
The average value of $H_{\rm 0}$ based on these time-delays measurements
clusters at about $70~{\rm km~s^{-1}~Mpc^{-1}}$ with 
an uncertainty of $\sim$ 10\%.
Thus, the estimate of $H_{\rm 0}$ obtained 
with the gravitational lensing method is consistent 
with the value obtained by the HST Key Project (HSTKP) on 
the Extragalactic Distance Scale.

\begin{figure}[htbp]
\plotone{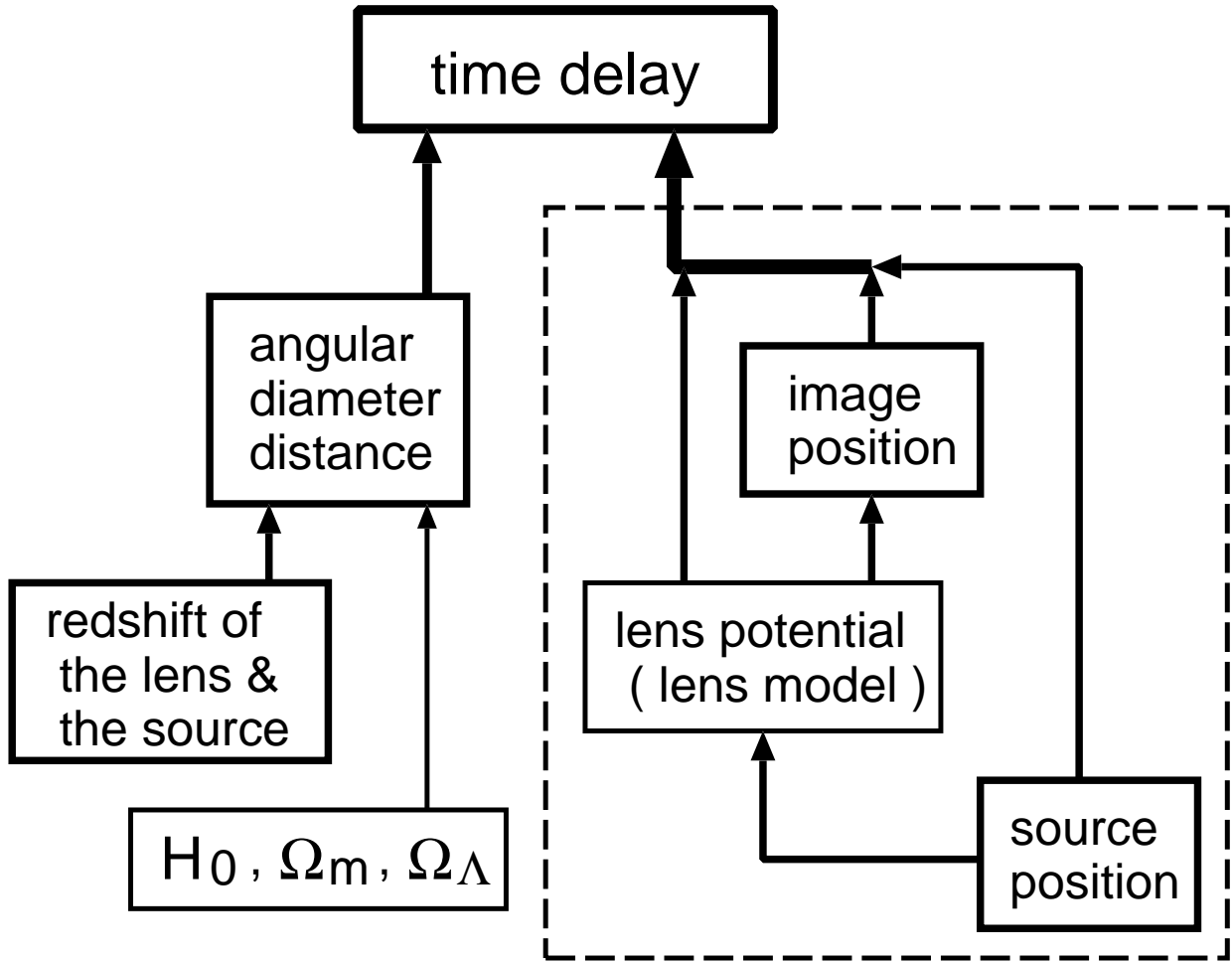}
\caption{A schematic diagram describing the calculation of the time delay 
between images of a gravitationally lensed system.}
\label{fig:schemview}
\end{figure}

Recently, WMAP first year result (Spergel et al. 2003) presents 
fundamental numbers of universe including $H_{\rm 0}$ with 
accuarcies better than various independent technique. 
Since the value of $H_{\rm 0}$ have been also measured with good accuracy, 
one may ask the question of whether there is a need to continue 
with measurements of time delays in lensed systems.
We point out that time delay measurements can provide
information in addition to $H_{\rm 0}$.
As shown in figure~\ref{fig:schemview}, 
the observed time delays depend not only $H_{\rm 0}$ 
but also on the cosmology, astrophysics and astrometry.
Even before the detection of the first gravitational lens system 
Refsdal (1966) had noted that the cosmological parameters 
($\Omega _{\rm m}$ and $\Omega _{\Lambda}$) can be constrained 
from the time delays 
between multiple images of a gravitationally lensed source. 
Recently, the possibility of probing the mass profiles of the lens objects 
was discussed by several groups (e.g., Oguri et al. 2002), 
and a technique of constraining the physical properties of the source objects, 
i.e., quasars, was investigated by Yonehara (1999).

The realization of the above possibilities may depend on 
the accuracy of the time delay measurements and 
the timescale of flare-like flux variations in quasars.
The former will be improved with time, though the latter is somewhat unclear.
Recently, Chartas et al. (2001), Dai et al. (2003) and Chartas et al. (2002) 
detected with the Chandra and XMM-Newton X-ray Observatories
rapid X-ray flares with durations of $\sim 1~{\rm ksec}$ 
in the gravitationally lensed quasars RX~J0911.4+0551 ($z = 2.8$), 
PG~1115+080 ($z = 1.72$) and APM~08279+5255 ( $z= 3.91$).
The presence of such rapid flares are exciting and suggests that, 
such rapid X-ray flares with the Kepler rotation period 
at the Schwarzschild radius do exist in quasars.
We anticipate that in the near future the number of detected quasars that show 
such rapid flux variations will increase.

In this paper, we describe a new promising application of 
gravitational lensing: we present a method of mapping the locations 
of flares in quasars employing an interferometric technique.
We expect that these goals will be attainable in the near future as
high precision measurements of the time delay between multiple images of 
gravitationally lensed quasars become available.
In section 2, we introduce a method to constrain the locations of 
flares in quasars which we refer to as 
the Gravitational Lens Interferometric Mapping Method (GLIMM), 
and show several applications of this method in section 3.
Finally, section 4 is devoted to a discussion of our proposed method.

\section{METHODOLOGY TO MAP THE LOCATION OF FLARES IN QUASARS}

In the following section we outline a method that allows one 
to extract  information about the location of flares in quasars 
from the observed time delay between images of a lensed quasar

The derivation of $H_{\rm 0}$ from measured time delays, in general,
assumes that the source is a point source, which is a reasonable
assumption for this derivation. 
However, astrophysical objects including quasar accretion disks 
have finite sizes, and the time delays due to gravitational lensing 
will depend on the location of the flares on the surface of the disks.
Since flares may occur at various locations on the disk 
(Kawaguchi et al. 1998, 2000), we expect 
the observed time delay to vary depending on where the flare originated.
This dependence of the time delay with the flare location is 
the basic idea of this paper and 
was initially investigated by Yonehara (1999).
Even in the case when X-ray flares are 
originated from shocks within a relativistic jet
(e.g., Rees 1978, Spada et al. 2001, Sikora et al. 2001), 
in principle, our proposing method is also applicable
\footnote{This is the case for multiple image blazars. 
 In such a situation, we may be able to probe the spatial structures 
 of relativistic jet in active galactic nuclei.}.
Therefore, in the case of shocks within a relativistic jet, 
it will be possible to map the locations of shocks within the jet.

Following, we briefly review the derivation of the time delay, 
describe the dependence of the time delay on the finite size of the source, 
and introduce the gravitational lens interferometric mapping method (GLIMM) 
that maps the locations of accretion-disk flares.

\subsection{Time delay and Time delay difference}

It is well known that the time delay ($\tau _{i,j}$) between 
multiple images $i$ and $j$ produced by the gravitational lensing effect 
is given by the following equation (e.g., Schneider, Ehlers, \& Falco 1992) :
\begin{equation}
 \tau _{i,j} = \frac{1}{2c} \frac{D_{\rm ol} D_{\rm os}}{D_{\rm ls}}
 \left( 1+z_{\rm l} \right) \left[ \Bigl\{ \left( \vec{\theta _i} - 
 \vec{\beta} \right) ^2 - \left( \vec{\theta _j} - \vec{\beta} 
 \right) ^2 \Bigr\} - 2 \Bigl\{ \psi ( \vec{ \theta _i } ) -
 \psi ( \vec{ \theta _j } ) \Bigr\}  \right], 
\label{eq:delayform}
\end{equation}
where $c$ is the speed of light, $D_{\rm ol}$, $D_{\rm os}$, and $D_{\rm ls}$
are the angular diameter distances (e.g., Weinberg 1972) 
between observer and lens, observer and source, 
and lens and source, respectively, 
$z_{\rm l}$ is the redshift of the lens, 
$\vec{\theta _i}$ and $\vec{\beta}$ are the angular positions of 
the $i$-th image and the source, respectively.
$\psi ( \vec{ \theta _i } )$ is the lens potential at $\vec{ \theta _i }$.
This lens potential may include external shear and/or contributions
of neighboring galaxies, depending on the nature of the lens.
A schematic diagram describing the calculation of the time delay 
is shown in figure~\ref{fig:schemview}.
For our purpose, we also derive the time delay 
between the same multiple images 
for different positions of the source, $\tau _{i,j}^{\prime}$.
Similar to equation~\ref{eq:delayform}, such a time delay is written as 
\begin{eqnarray}
 \tau _{i,j}^{\prime} &=& \frac{1}{2c} \frac{D_{\rm ol} D_{\rm os}}{D_{\rm ls}}
 \left( 1+z_{\rm l} \right) \left[ \Bigl\{ \left( \vec{\theta _i} 
 + d \vec{\theta_i} - \vec{\beta} - d \vec{\beta} \right) ^2 - 
 \left( \vec{\theta _j} + d \vec{\theta _j} - \vec{\beta} - d \vec{\beta} 
 \right) ^2 \Bigr\} \right. \nonumber \\
 & & \left. - 2 \Bigl\{ \psi ( \vec{ \theta _i} + d \vec{\theta _i} ) 
 - \psi ( \vec{ \theta _j } + d \vec{\theta _j}) \Bigr\}  \right], 
\label{eq:delaydash}
\end{eqnarray}
where $d \vec{\beta}$ is the difference of the source position from 
$\vec{\beta}$ in the equation~\ref{eq:delayform}, and 
$d \vec{\theta _i}$ is the corresponding deviation of the image position.

\begin{figure}[htbp]
\plotone{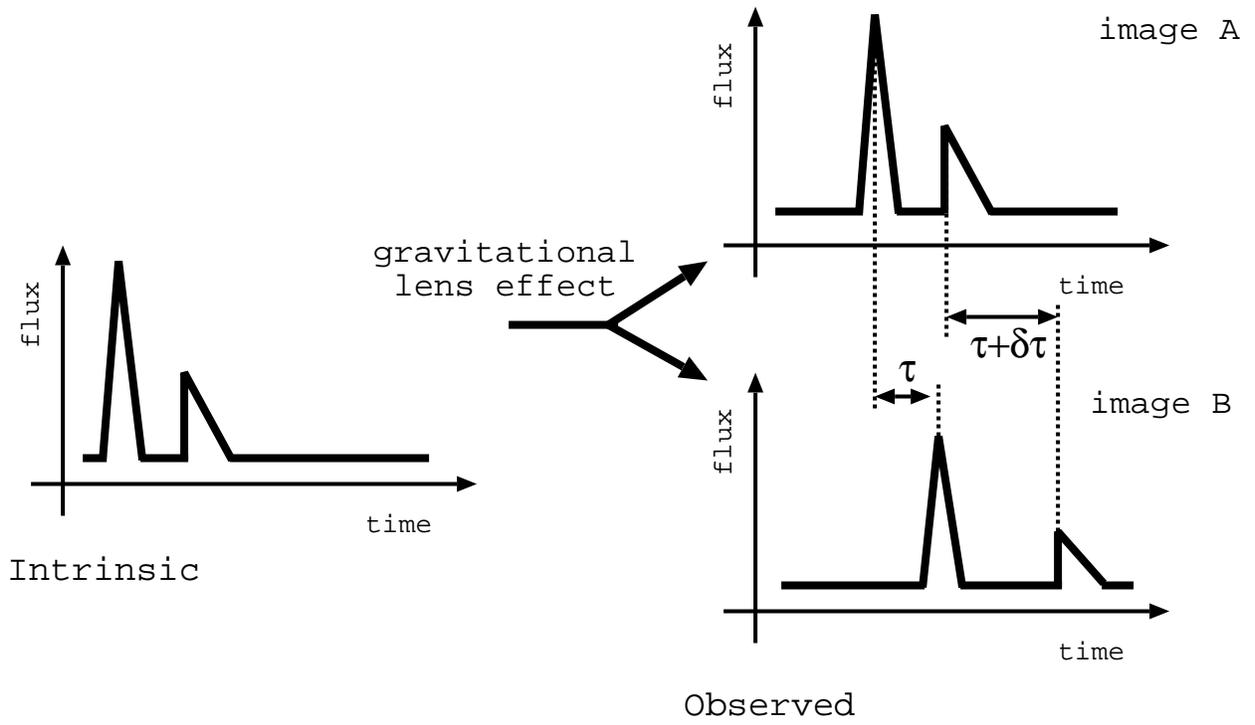}
\caption{Schematic plots of the difference in the time delay between different 
flares in the same image pair of a lensed quasar.}
\label{fig:schcurve}
\end{figure}

Given the source position and the lens potential 
in a gravitational lens system,
one can obtain the ideal image positions by solving the lens equation.
Thus, $\vec{\theta _i}$ and $d \vec{\theta _i}$ are
determined by $\vec{\beta}$ and $d \vec{\beta}$ 
through the  ``lens equation''
\footnote{Practically, we do not know the source position and 
the lens potential but know the image positions. Thus, using the observed
image positions and modeling a lens potential, one can calculate
the corresponding source position via the lens equation.
After that, one can estimate the expected time delays between images.}.  
Subtracting equation~\ref{eq:delayform} ($\tau _{i,j}$) from 
equation~\ref{eq:delaydash} ($\tau _{i,j}^{\prime}$),  
we can obtain the time delay difference 
(figure~\ref{fig:schcurve}) between images, 
$\delta \tau = \tau _{i,j}^{\prime} - \tau _{i,j}$, 
as a function of the difference between the source positions, $d \vec{\beta}$.

If the difference in the source positions is small enough 
compared with the typical lens size, the latter time delay can be 
expressed in the form of a simple Taylor expansion.
After some algebra, we find that the time delay difference ($\delta \tau$) 
can be expressed as,
\begin{equation}
\delta \tau \simeq 
 \frac{(1 + z_{\rm l})}{c} \frac{D_{\rm os}D_{\rm ol}}{D_{\rm ls}}
 \left[ \left( \vec{\theta _i} - \vec{\theta _j} \right)
 \cdot d \vec{\beta} \right] , 
\label{eq:beforeglimm}
\end{equation}
where, $d \vec{\beta}$ represents the angular separation of 
two slightly different source positions.
If the physical length that corresponds to this angular separation is 
$\delta \vec{r}$, then $d \vec{\beta}$ can be rewritten as 
$\delta \vec{r} / D_{\rm os}$ and equation~\ref{eq:beforeglimm} 
can be rewritten in the form, 
\begin{equation}
\delta \tau \simeq 
 \frac{(1 + z_{\rm l})}{c} \frac{D_{\rm ol}}{D_{\rm ls}}
 \left[ \left( \vec{\theta_i} - \vec{\theta_j} \right)
 \cdot \delta \vec{r} \right] .
\label{eq:glimmbasis}
\end{equation}

\begin{figure}[htbp]
\plotone{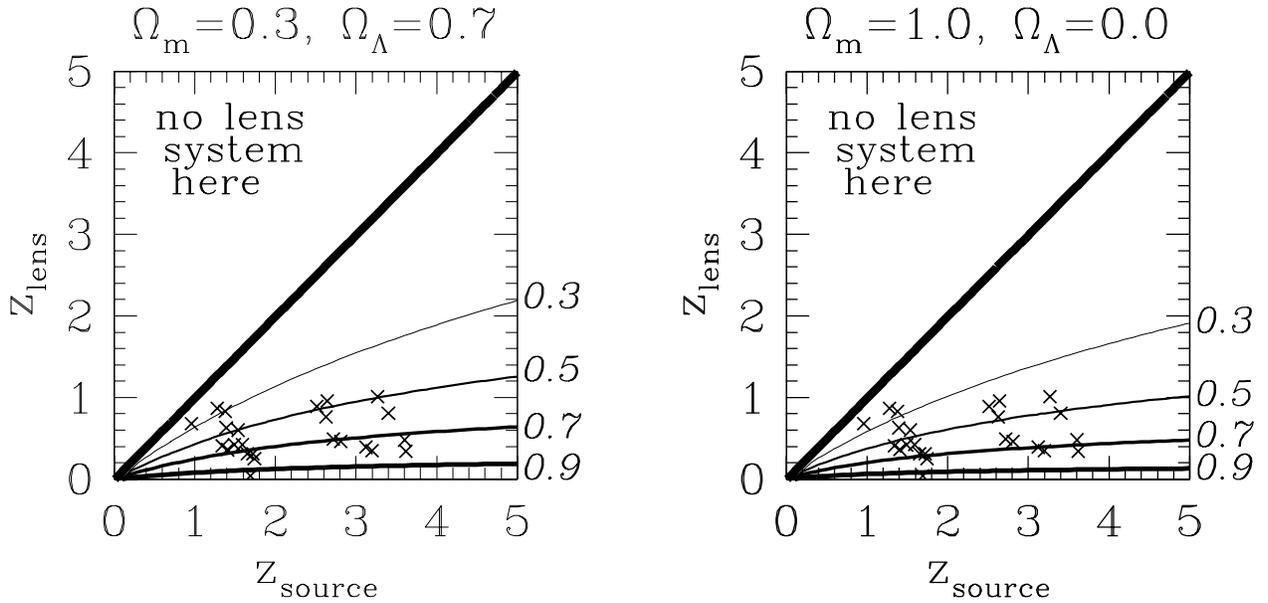}
\caption{The source and the lens redshift dependence of 
$D_{\rm ls} / D_{\rm ol}$ (not $D_{\rm ol} / D_{\rm ls}$) 
is shown in this figure. 
From thick to thin lines (or from bottom to top), contours of 
$D_{\rm ls} / D_{\rm ol} =0.9$, $0.7$, $0.5$ and $0.3$ are presented. 
Currently known gravitationally lensed quasars are also presented 
with crosses.
Cosmological parameters in the left and the right panel are
 $(\Omega _{\rm m}, \Omega _{\Lambda}) = (0.3, 0.7)$ and 
 $(\Omega _{\rm m}, \Omega _{\Lambda}) = (1.0, 0.0)$, respectively.}
\label{fig:zddep}
\end{figure}

Equation~\ref{eq:glimmbasis} provides a relation between the location of 
flares and the difference in time delays produced by these flares.
A useful aspect of this equation is that it contains no dependence 
on the lens model nor on $H_{\rm 0}$ and depends only slightly on
$\Omega _{\rm m}$ and $\Omega _{\Lambda}$ through the ratio 
$D_{\rm ol} / D_{\rm ls}$ (or $D_{\rm ls} / D_{\rm ol}$).
The value of $D_{\rm ol}/ D_{\rm ls}$ is of order unity and has 
a weak dependences on the redshifts of the lens and the source 
(or lens systems), but we can calculate 
$D_{\rm ol} / D_{\rm ls}$ for any given redshifts 
as shown in figure~\ref{fig:zddep} 
\footnote{This figure shows the contour of $D_{\rm ls} / D_{\rm ol}$}. 
This equation consists of two simple observables; 
the time delay difference and image positions,
and the quantity that we intend to explore, the separation between flares.
Application of equation~\ref{eq:glimmbasis}  provides 
the locations of flares on the quasar accretion disk 
from two simple observables without any particular assumptions.

\subsection{Gravitational Lens Interferometric Mapping Method (GLIMM)}

\begin{figure}[htbp]
\plotone{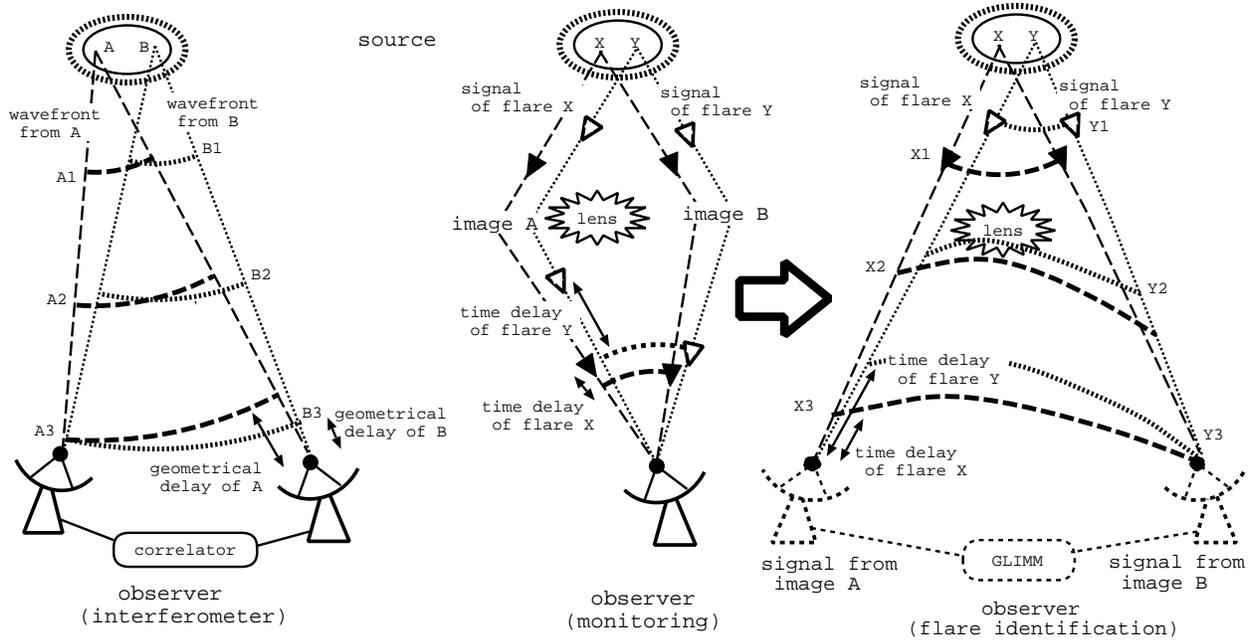}
\caption{Schematic figures of the gravitational lens interferometric Mapping 
Method (GLIMM) compared to a radio interferometric observation.}
\label{fig:schinter}
\end{figure}

Next, we describe in more detail a method for mapping accretion disk flares.
A schematic diagram of the Gravitational Lens Interferometric Mapping Method 
(hereafter referred to as GLIMM) is presented in figure~\ref{fig:schinter}. 
 
We first rewrite equation~\ref{eq:glimmbasis} in terms of typical values for 
the image separations and the expected difference in the time delays, 
\begin{equation}
\bigl| \delta \vec{r} \bigr|
 \sim 10 r_{\rm g} \left( \frac{M_{\rm BH}}{10^9 M_{\odot}} \right) ^{-1}
 \left( \frac{\delta \tau}{1~{\rm sec}} \right)
 \left( \frac{1 + z_{\rm l}}{2} \right) ^{-1} 
 \left( \frac{D_{\rm ls}}{D_{\rm ol}} \right)
 \left( \frac{\bigl| \vec{\theta_i} - \vec{\theta_j} \bigr| \cdot \cos \phi}
 {1~{\rm arcsec}} \right) ^{-1}, 
\label{eq:sizeest}
\end{equation}
or
\begin{equation}
\bigl| \delta \vec{r} \bigr| 
 \sim 200 \left( \frac{\delta \tau}{1~{\rm sec}} \right)
 \left( \frac{1 + z_{\rm l}}{2} \right) ^{-1} 
 \left( \frac{D_{\rm ls}}{D_{\rm ol}} \right)
 \left( \frac{\bigl| \vec{\theta_i} - \vec{\theta_j} \bigr| \cdot \cos \phi}
 {1~{\rm arcsec}} \right) ^{-1} 
 ~({\rm AU}),
\label{eq:sizeau} 
\end{equation}
where, $\phi$ is the angle measured from the direction between 
images $i$ and $j$, and  $r_{\rm g}$ is the Schwarzschild radius of 
the central black hole with mass $M_{\rm BH}$.
A simple application of these equations indicates that
a measurement of the difference in the time delays of two flares to 
an accuracy of $\sim 1~{\rm sec}$ results in a value for the separation
of these flares that is accurate to about $10 r_{\rm g}$ for a quasar 
containing a $10^9 M_{\odot}$ black hole.

\begin{figure}[htbp]
\plotone{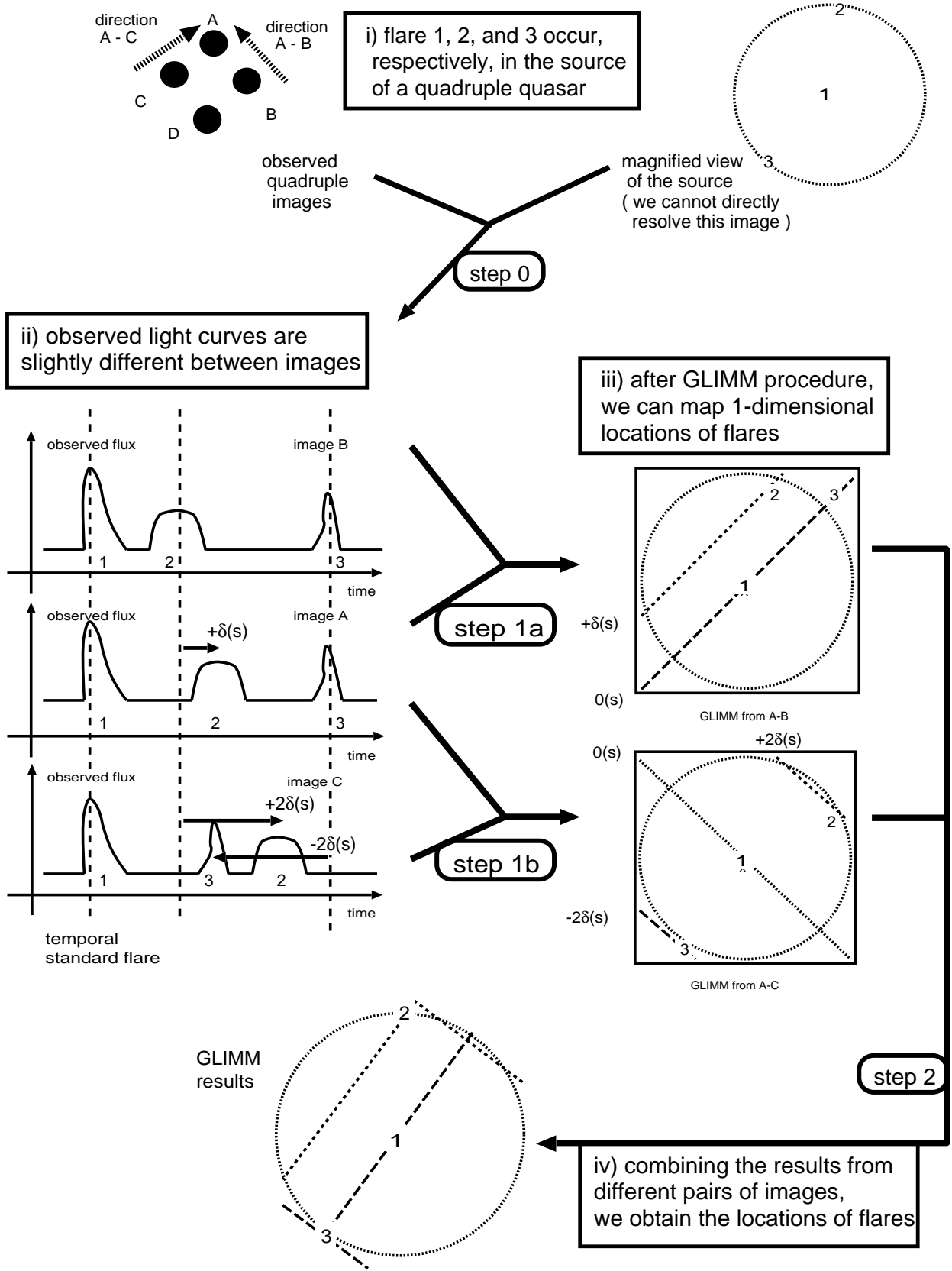}
\caption{A flow diagram illustrating the use of the Gravitational Lens 
Interferometric Mapping Method (GLIMM).}
\label{fig:glimm}
\end{figure}

When equation ~\ref{eq:glimmbasis} or~\ref{eq:sizeest} is
applied to a gravitationally lensed quasar with more than three images,
it provides tighter constraints on the separation of flares than 
when applied to gravitational lens systems with fewer images.
For gravitational lens systems with only two images,
the time delay difference $\delta \tau$ is most sensitive to flares separated 
along the direction connecting the two images and is not sensitive to 
flares that are separated orthogonal to the image direction.
For a gravitational lens system with more than three images, the time delay
difference $\delta \tau$ is sensitive to flares separated in any direction
allowing for two-dimensional mapping of the locations of the flares.
A flow diagram illustrating the use of GLIMM is 
presented in figure~\ref{fig:glimm}.

\section{A Specific Example: Case of RX~J0911.4+0551}

Our earlier description of using GLIMM to infer two-dimensional distributions 
of flares in accretion disks was somewhat qualitative and, 
therefore, in the following subsection, we present numerical simulations to 
confirm our estimations and test the feasibility of 
our proposed mapping method. 
For our purpose, we have chosen gravitational lens systems with more than 
three images including the quadruple system, RX~J0911.4+0551, 
in which Chartas et al. (2001) detected a rapid X-ray flare.

\subsection{Lens Model and Time Delay Contour}

\begin{figure}[htbp]
\plotone{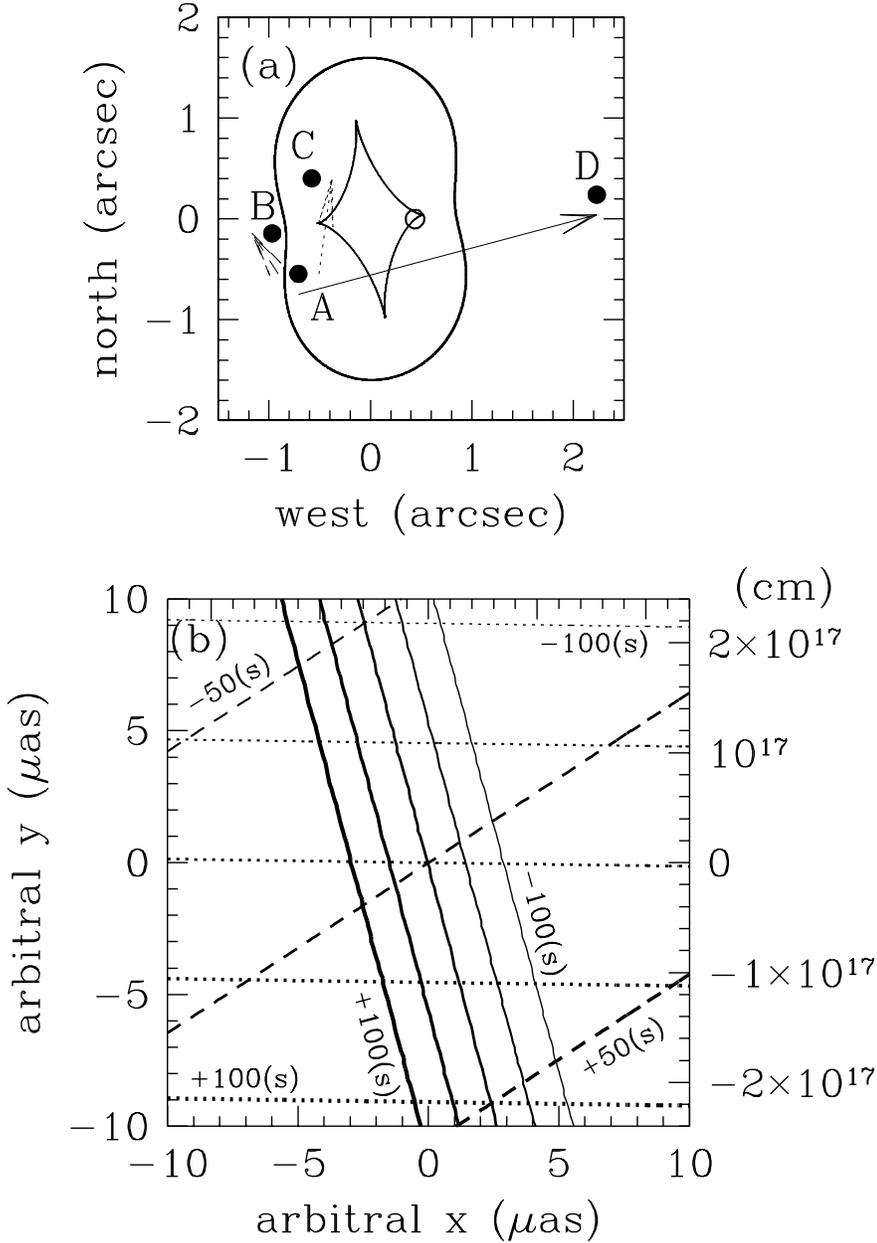}
\caption{In the upper panel (a), we show the caustic 
(inner, thin and cuspy curve), critical curve (outer, thick and smooth curve), 
source position (open circle), and image positions (filled circles) resulting 
from a fit of a lens model to the gravitational lens system RX~J0911.4+0551. 
We model the potential of the lens with a pseudo-isothermal elliptic mass 
distribution (PIEMD) lens model plus shear. 
The direction vectors connecting image pairs are represented by arrows. 
These directions correspond to mapping directions of GLIMM.
Lower panel (b) shows contours of time delay differences between 
images A and B (dashed lines), images A and C (dotted lines), 
and images A and D (solid lines). 
Along the direction of arrows shown in the upper panel, 
or from thick to thin line, values of time delay difference is 
$+100$, $+50$, $\pm 0$, $-50$, and $-100$ (sec), respectively.
``$+$ ($-$)'' means that the time delay on this location 
is longer (shorter) than that on the center.
In the case of the differece between image A and B, or dashed lines, 
$+100$ and $-100$ (sec) contours are out of range of this figure.}
\label{fig:contour}
\end{figure}

We model the potential of the lens in the gravitational lens system 
RX~J0911.4+0551 with a pseudo-isothermal elliptic mass distribution (PIEMD) 
lens model (e.g., Kassiola \& Kovner 1993) plus external shear.
The observable constraints are the positions of the lensed images 
and lensing galaxy.
Observed flux ratios were not considered in our lens model since 
these may be affected by many factors, such as  
differential extinction and microlensing.
By fitting our lens model using a $\chi ^2$ minimization method
(e.g., Kayser et al. 1990; Kochanek 1991), 
we obtain the best fit parameters which are presented in 
table~\ref{tab:posrxj} and table~\ref{tab:prmrxj}.
The reduced $\chi ^2$ is less than $1$ and the best fit lens model 
reproduces the observed positions of the images accurately.
The resulting caustic, critical curve, source position, and 
image positions are depicted in the upper panel of figure~\ref{fig:contour}. 
Using these best fit lens parameters, 
we can calculate the time delay difference between images.
Contours of the time delay difference are presented 
in the lower panel of figure~\ref{fig:contour}. 
As expected from equation~\ref{eq:glimmbasis}, 
the contour lines are perpendicular to the direction vector connecting 
image pairs which are represented by arrows in the upper panel 
of figure~\ref{fig:contour}.
It is also apparent from equation~\ref{eq:glimmbasis} that
the density of the contour levels of the time delay differences is 
proportional to the image separation.
To convert from apparent angular distances to actual distances, 
we have assumed $H_{\rm 0} = 70~{\rm km~s^{-1}~Mpc^{-1}}$,
$\Omega _{\rm m} = 0.3$, and $\Omega _{\Lambda} = 0.7$.
Adopting this cosmology and considering an accretion disk with a size of 
$1000 r_g$ extending around a $10^9 M_{\odot}$ black hole
\footnote{It will have a radius of $\sim 2 \times 10^{17}~{\rm cm}$},  
the expected time delay difference between images A and D varies from 
$\sim -200~{\rm sec}$ to $\sim +200~{\rm sec}$ for source positions 
spreading across the assumed extent of the accretion disk.
This result supports our analytic estimate in the previous subsection, 
since in the case of RX~J0911.4-0551 ($z_{\rm ol} \sim 0.769$, 
$z_{\rm os} \sim 2.8$, 
$D_{\rm ls} / D_{\rm ol} \sim 0.5$ (see figure~\ref{fig:zddep}), 
and $\theta_{\rm A} - \theta _{\rm D} \sim 3.0~{\rm arcsec}$),  
the expected variation of the time delay difference between images A and D 
for the assumed extent of the accretion disk ($\sim 10^{17} {\rm (cm)}$) 
becomes $\sim \pm 200~{\rm sec}$ from equation~\ref{eq:sizeest}.
These simulations also indicate that the spatial resolution of mapping flares 
using GLIMM may reach several $r_{\rm g}$ when observing techniques improve  
to the level of constraining time delays to an accuracy of $\sim 1~{\rm sec}$.

\subsection{Simulated Light Curves}

\begin{figure}[htbp]
\plotone{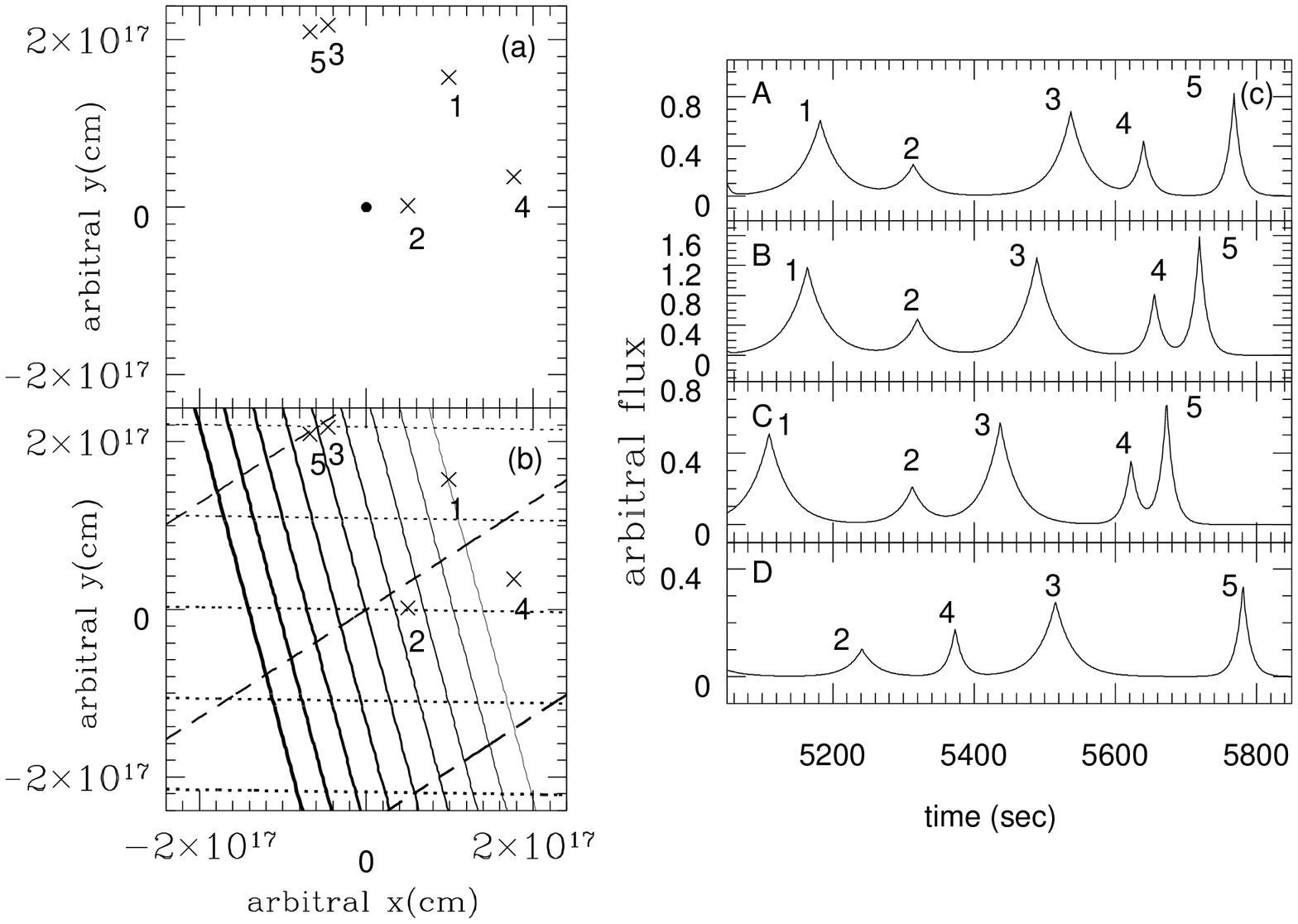}
\caption{In the upper left panel (a), we show the locations (``$\times$'') 
of five flares on the source plane.
The locations of five flares are assigned a number ranging from 1 to 5.
The center of this figure is denoted as dot.
In the lower left panel (b), 
we shows the contours of the time delay differences.
We shifted the light curves of images B, C and D by the median time delay 
between these three images and image A. 
We then applied GLIMM to the simulated light curves shown in the right panel 
(c) to determine the locations of the flares. 
The inferred locations of the flares are over-plotted with $\times$  
in the lower left panel (b). 
The meaning of the lines shown in the lower left panel is similar 
to that presented in figure~\ref{fig:contour}.}
\label{fig:locatflr}
\end{figure}

To simulate light curves of flares in gravitational lensed quasars, 
we take the following steps; 
(1) We start the simulation by creating a flare with an exponential growth 
and decay profile. The flare is produced randomly in time and space on 
the surface of the accretion disk. The intensity of the flare is also 
chosen to be random.
(2) We numerically calculate the time delay and fluxes of the flare 
for each image.
(3) Steps 1 and 2 are repeated a number of times to produce a light curve 
for each image. 
In figure~\ref{fig:locatflr}(a), we show the simulated location 
of five flares on the surface of the disk.
In figure~\ref{fig:locatflr}(b), we show the contours of the time delay 
difference. The simulated light curves of the flares in the four images 
are shown in figure~\ref{fig:locatflr}(c). 
We shifted the light curves of images B, C, and D by the median time delay 
between these three images and image A.
We then applied GLIMM to the simulated light curves to determine 
the locations of the flares. 
The inferred locations of the flares are over-plotted with 
filled circles in figure~\ref{fig:locatflr}(b). 

It is possible to identify flares of common origin in the light curves 
of images since each flare has a unique shape and the duration of a flare 
is not affected by the gravitational lens effect.
Variations of the flux ratios of flares between images are expected 
to occur because of the different locations of the flares.
For our simulation, the variations in flux ratios were estimated to be less 
than 1 \% and such a tiny change of flux ratios does not complicate 
the identification of the flares.
Therefore, if we take into account flux ratios between images, 
we will successfully identify the corresponding flares between images. 

As shown by observational (e.g., Giveon et al., 1999, 
Uttley, McHardy, \& Papadakis 2002) and 
theoretical (e.g., Kawaguchi et al. 1998, 2000) studies 
for flux variabilities of quasar (or active galactic nuclei), 
flares with small amplitude occur randomly and frequently.
Thus, unfortunately, flares crowd upon each other 
and we may not be able to apply GLIMM to such flares. 
In contrast, the same studies also shows that 
flares with large amplitude do not occur so frequently, 
and we can separate individual flares.
Moreover, flares with large amplitude are known to show 
stochastic natures of its shape, amplitude and duration 
by the previous observational and theoretical studies
\footnote{In the case of flares with small amplitude, 
we cannot say much things about natures of their shape 
due to the limited time resolution or effective area of observational 
instruments, and the limited resolution for numerical simulations. }. 
Taking into account such properties, individual flare should have unique shape 
and it can be quite rare case or unrealistic that 
the almost identical flares with different origin occur.
Thus, we can successfully identify the corresponding flares 
between multiple images at least in the case of large flares.

In table~\ref{tab:expdelay}, we list the time delay difference of 
common origin flares between images.  
The median time delays between images have been subtracted from 
the time delay differences presented in table~\ref{tab:expdelay}.
These time delay differences are expressed with respect to the times 
of the flares that are detected in image A.
Measured time delay differences can be used to constrain the locations 
of the flares using GLIMM. We apply GLIMM to the two pairs of images 
in our simulated case of a quadruple lens system. 
The non-parallel directions of these image pairs allows us to produce 
a 2-dimensional map of the flare. 
We apply GLIMM to the remaining image pair to cross-check our results 
with those obtained from the application of GLIMM to the other two image pairs.

\section{DISCUSSIONS}

In this paper, we have presented a new application of gravitational lensing 
that may be employed in the near future; 
we have described how measurements of the time delays of accretion disk 
flares with accuracies of a few seconds may provide 2-dimensional mapping of 
the flare positions.  
At present, time delay accuracies of order  
$1 {\rm (sec)}$ seem to be somewhat unrealistic.
However, we anticipate that such accuracies will become attainable 
with future satellites.

\subsection{Another Possible Application of GLIMM}

Flares may also originate from supernovae 
in the host galaxy of quasars or starburst galaxies.
If the total extent of detected flares are much larger 
than expected in quasar accretion disks, 
flare mapping can be realized by coordinating 
current observational instruments in a proper way 
such as the operation of the world wide monitoring networks 
(e.g., GLITP \footnote{{\tt http://www.iac.es/proyect/gravs\symbol{"5F}lens/GLITP/index.html}}). 
Thus, even with the currently available time resolution, 
GLIMM will be able to constrain the location of
supernovae in the host galaxy of quasars or starburst galaxies.

Currently, there is a considerable amount of effort being devoted to 
the investigation of a possible physical connection between quasars and 
starburst activity (e.g., Boyle \& Terlevich 1998, Ohsuga et al. 1999). 
We expect that the host galaxies of several lensed quasars may 
exhibit starburst activity and supernovae explosions may 
occur frequently in these galaxies. 
The total extent of a starburst region is expected to be $\ge 100~{\rm pc}$.
Using equation~\ref{eq:glimmbasis} or ~\ref{eq:sizeau},  
the expected time delay difference between different supernovae is predicted
to be of the order of days due to the large spatial separation between 
supernovae in the host galaxy.  
Recently, Goicoechea (2002) reported the presence of multiple time delays 
in the double quasar, Q0957+561, and attributed this to the large extent of 
the variable emission region. They suggested that this variability 
is partly caused by some explosive phenomena of stellar objects 
in the host galaxy. 
Other proposed mechanisms of quasar variability are 
instabilities of the accretion disk 
based on the statistical properties of the flares (e.g, Kawaguchi et al. 1998).
If it is shown that quasar variability is partly produced by supernova type 
explosions, we can apply GLIMM in these cases to map their 
spatial distribution and thus map the starburst region and/or 
locations of supernova explosions in distant galaxies.
Such information may provide insight to the star formation history 
or evolution process of (starburst) galaxies.
  
We also note that if the conclusions of Goicoechea (2002) are correct,
we can plan a detailed spectroscopic observation of a supernova 
by triggering on an event detected in the leading image 
and wait by the known time delay to make a detailed spectroscopic observation
of the delayed event in another lensed image.
This will enable us to obtain a spectrum of the very early stage of a 
supernova.
We conclude that  GLIMM can be a powerful tool not only for 
understanding accretion disk physics, but also for stellar 
and/or supernova physics.

\subsection{Practical Difficulties}

Here, we discuss practical difficulties that may arise when applying
GLIMM to real data. To achieve our goal of mapping flares 
in quasar accretion disks, 
satellites with collecting areas larger than those of present X-ray telescopes
will be required. As indicated in section 2, 
mapping of accretion disk flares via GLIMM requires 
measuring time delays of flares to accuracies of a few seconds. 
Recently, a rapid X-ray flare with a duration of about $\sim 2000~{\rm sec}$ 
was detected in a Chandra observation of the gravitationally lensed quasar 
RX~J0991.4+0551 (Chartas et al. 2001).
X-ray flares have also been detected with Chandra in several other 
gravitational lens systems such as PG~1115+080 and APM~08279+0552 
(Dai et al. 2003, Chartas et al. 2002). 
If these flares are intrinsic, i.e., originate from the quasar accretion disk, 
GLIMM can, in principle, be applied to map 
the spatial structure of the quasar accretion disk. 
For example, the flare that Chartas et al. (2001) detected in 
RX~J0991.4+0551 shows $30$ counts in $2 {\rm (ksec)}$, and 
the relatively low effective area of present X-ray telescopes 
precludes time delay measurements with the specific accuracies 
needed to successfully apply GLIMM. 
However, future X-ray missions with large effective areas such as XEUS 
\footnote{{\tt http://astro.esa.int/SA-general/Projects/XEUS/}} 
may be capable of measuring the arrival times of flares 
to accuracies of a few seconds.
In the case of XEUS, effective areas will be increased 
by 3 orders of magnitude, and expected photon counts 
will be $30 \times 10^3 = 3 \times 10^4$ in $2 {\rm (ksec)}$ or 
$15 {\rm (c/s)}$ for the flare that Chartas et al. (2001) detected.
Assuming a low background, a signal-to-noise ratio 
within a $1 {\rm (sec)}$-bin 
becomes $\sim 4$, and this value is sufficient for a 3-sigma detection. 

One practical difficulty in applying GLIMM is the identifications 
of flares of common origin.
If the actual flare is isolated, such an identification will be simple
because the durations of flares are not affected 
by the gravitational lens effect 
and flux ratios are almost identical over the quasar accretion disk.
On the other hand, in a worst case scenario, a flare could be superposed 
on another flare,  
and the identification of a flare may not be an easy task in this situation.
If the typical time separations between flares are comparable or smaller than
the typical time scales of flares, a superposition of flares may occur.
In such a case, the fading and/or brightening part of a flare shape 
will be strongly distorted, and its original shape will be altered.
However, the identifications of the peaks of flares may not be 
so difficult to determine.
After we determine the peaks of superposed flares, we can subtract
individual shapes of flares from observed light curves
by extrapolating the light curves from the peak, 
assuming exponential growth and decay of the light curves.
If there is a particularly active region, unfortunately,  
the situation will become worse.
The reason is that such region could possibly produce 
multiple flares of very similar types, and identifications of 
corresponding flares can be difficult.
To overcome this problem, some statistical arguments will be required, 
but there is no guarantee to clearly extract scientific results 
from such multiple flares.

Another practical difficulty in using GLIMM is that 
other confusing phenomena may be superposed 
on the light curves of the quasar, 
e.g., quasar microlensing and additional gravitational lens effects 
due to the small scale structures in the lens galaxy.
Flux variability induced by quasar microlensing is distinguishable from 
intrinsic variability of quasars because quasar microlensing events 
do not lead to time delayed variations.  
Furthermore, typical timescales of quasar microlensing events are long 
(e.g., Wambsganss et al. 1990) and can be effectively removed 
prior to applying GLIMM.   
Additional gravitational lens effects due to small scale structures
in the lens galaxy may change the time delay difference.
However, even if CDM substructures are present, 
such events will lead to changes of the time delay 
of order of a few percent (Yonehara et al. 2003). 
Moreover, even if CDM substructure is present along the line of sight
toward an image, we can calibrate the remaining image pairs 
in multiple image quasars with 4 or more images. 
This calibration may allow us to probe the existence of 
small scale structures in the lens galaxy.

Finally, we briefly mention about gravitationally lensed quasars 
with less than three images.
In section 2, we showed that applying GLIMM to gravitational lens systems 
with three or more images may provide 2-dimensional mapping of 
accretion disk flares. 
We note that GLIMM can also be applied to double image quasars resulting in 
1-dimensional mapping of the spatial distribution of flares 
on the accretion disk.

Current X-ray telescopes do not have the sensitivities to 
allow us for the implementation of GLIMM, even in the case of 
the most luminous gravitationally lensed quasars. 
But this technique will work 
with better instruments developed in near future.
Until now, the lensing method has been used primarily to obtain 
a direct value of $H_{\rm 0}$.
We hope that in the near future, the lensing method may 
allow us to map the spatial distribution of flares 
in accretion disks of distant quasars.

\acknowledgements

The authors would like to express their thanks to 
J.A. Mu\~noz, J. Wambsgan\ss, E. Falco, R. Schmidt, M. Umemura, 
H. Susa, K.Mitsuda, for their valuable comments. 
Furthermore, the authors would like to thank
the anonymous referee for important comments and suggestions. 
This work was supported in part by the Japan Society for
the Promotion of Science (09514).

\clearpage

\begin{deluxetable}{lccccc}
\tabletypesize{\scriptsize}
\tablecaption{Astrometry for RX~J0911.4+0551.\label{tab:posrxj}}
\tablewidth{0pt}
\tablehead{
\colhead{position} & \colhead{image A} & \colhead{image B} & 
\colhead{image C} & \colhead{image D} & \colhead{galaxy}
}
\startdata
$x$ (arcsec) & 0.000$\pm$0.004 & -0.259$\pm$0.007 & +0.013$\pm$0.008 & 
 +2.935$\pm$0.002 & +0.709$\pm$0.026  \\
$y$ (arcsec) & 0.000$\pm$0.008 & +0.402$\pm$0.006 & +0.946$\pm$0.008 &
 +0.785$\pm$0.003 & +0.507$\pm$0.046  \\ 
\enddata

\tablecomments{Positions of images and lensing galaxy relative to 
 image A (the brightest image), were taken from Burud et al. (1998). 
$x$ and $y$ positions are defined positive toward north and west.}

\end{deluxetable}

\begin{deluxetable}{lc}
\tabletypesize{\scriptsize}
\tablecaption{Lens model parameters for RX~J0911.4+0551.\label{tab:prmrxj}}
\tablewidth{0pt}
\tablehead{
\colhead{Parameter} & \colhead{Value} 
}
\startdata
$\gamma $ & 0.3179 \\
$\phi _{\gamma} $ (degree)& 102.30 \\ 
$E_{\rm 0}$ (arcsec)& 1.1095  \\
$\epsilon$ & 0.0948 \\
$\omega$ (arcsec) & 0 (fix) \\ 
$\phi _{\rm lens}$ (degree) & 140.33 \\
$\beta _x$ (arcsec) & 0.4388 \\
$\beta _y$ (arcsec) & 0.0013 \\
\enddata
\tablecomments{Best fit parameters for RX~J0911.4+551. 
Specifically, we present the external shear, $\gamma$, 
the PIEMD parameters, $E_{\rm 0}$, $\epsilon$ and $\omega$, and 
the source positions $\beta _x$ and $\beta _y$.
The notation of the PIEMD parameters is similar to that used 
in Kassiola \& Kovner (1993). 
Position angles of the external shear, $\phi _{\gamma}$, and  
of the ellipsoid, $\phi _{\rm lens}$, are measured from north to east. 
Due to the absence of a fifth image in this lens system, 
we have set the core radius, $\omega$, to zero. 
The reduced $\chi ^2$ of this lens model is $0.65$.} 
\end{deluxetable}

\begin{deluxetable}{lrrrrr}
\tabletypesize{\scriptsize}
\tablecaption{Time delay difference of simulated flares.\label{tab:expdelay}}
\tablewidth{0pt}
\tablehead{
\colhead{image} & \colhead{flare 1} & \colhead{flare 2} & \colhead{flare 3} &
\colhead{flare 4} & \colhead{flare 5} 
}
\startdata
 A & 0 (fix) & 0 (fix) & 0 (fix) & 0 (fix) & 0 (fix) \\
 B & -18 & +6 & -48 & +15 & -39 \\
 C & -72 & -1 & -100 & -18 & -95 \\
 D & -211 & -73 & -22 & -267 & +13 \\ \hline
\enddata
\tablecomments{Time delay differences of flares shown 
in figure~\ref{fig:locatflr}.
The time delays have units of seconds.
These time delay differences are expressed with respect to the times 
when the flares are detected in image A. 
Median time delay between image A and other images have subtracted.}
\end{deluxetable}

\clearpage

\begin{thebibliography}{}

\bibitem[]{} Boyle, B.J., \& Terlevich, R.J. 1998, MNRAS, 293, L49

\bibitem[]{} Burud, I., et al. 1998, ApJ, 501, L5

\bibitem[]{} Chartas, G., et al. 2001, ApJ, 558, 119

\bibitem[]{} Chartas, G., et al. 2002, ApJ, 579, 169

\bibitem[]{} Dai, X., et al. 2003, in preparation

\bibitem[]{} Giveon, U., et al. 1999, MNRAS, 306, 637

\bibitem[]{} Goicoechea, L.J. 2002, MNRAS, 334, 905

\bibitem[]{} Kassiola, A., \& Kovner, I. 1993, ApJ, 417, 450

\bibitem[]{} Kawaguchi, T., et al. 1998, ApJ, 504, 671

\bibitem[]{} Kawaguchi, T., et al. 2000, PASJ, 52, L1

\bibitem[]{} Kayser, R., et al. 1990, ApJ, 364, 15

\bibitem[]{} Kochanek, C.S. 1991, ApJ, 373, 354

\bibitem[]{} Oguri, M., et al. 2002, ApJ, 568, 488

\bibitem[]{} Ohsuga, K., et al. 1999, PASJ, 51, 345

\bibitem[]{} Rees, M. 1978, MNRAS, 184, 61

\bibitem[]{} Refsdal, S. 1964, MNRAS, 128, 295

\bibitem[]{} Refsdal, S. 1966, MNRAS, 132, 101

\bibitem[]{} Schneider, P., Ehlers, J., \& Falco, E.E. 1992, Gravitational Lenses, 2nd ed. (New York:Springer-Verlag)

\bibitem[]{} Sikora, M., et al. 2001, ApJ, 561, 1154

\bibitem[]{} Spada, M., et al. 2001, MNRAS, 325, 1559

\bibitem[]{} Spergel, D.N., et al. 2003,  submitted to ApJ (astro-ph/0302209)

\bibitem[]{} Uttley, P., McHardy, I.M., \& Papadakis, I.E. 2002, MNRAS, 332, 231

\bibitem[]{} Walsh, D., Carswell, R.F., \& Weymann, R.J. 1979, Nat., 279, 381

\bibitem[]{} Wambsganss, J., Paczy\'nski, B. \& Schneider, P. 1990, ApJ, 358, L33

\bibitem[]{} Yonehara, A. 1999, ApJ, 519, L31

\bibitem[]{} Yonehara, A., Susa, H., \& Umemura, M. 2003, in preparation. 

\end{thebibliography}
\end{document}